\documentclass[12pt]{article}

\usepackage{etex}
\usepackage{fullpage}
\usepackage{setspace}
\usepackage{comment}
\usepackage[pdftex]{graphicx}
\usepackage{rotating}
\usepackage{amsmath}
\usepackage{amsthm}
\usepackage{amssymb}
\usepackage{graphicx}
\usepackage{fullpage}
\usepackage{setspace}
\usepackage{natbib}
\usepackage{color}
\usepackage{booktabs}

\usepackage{pgf}
\usepackage{pgfplots}
\usepackage{textpos}
\usepackage{graphicx}
\usepackage{pstricks}
\usepackage{framed}
\usepackage{varwidth}
\usepackage[normalem]{ulem}
\usepackage{xstring}

\usepackage{algorithm}
\usepackage{algpseudocode}

\newcommand{\OMIT}[1]{\relax}   
\def\text{{\rm}}

\def\Pr{P}

\def\E{\hbox{E}}

\newcommand{\bma}[1]{\mbox{\boldmath $#1$}}

\newcommand{\bA}{ {\bma{A}} }
\newcommand{\ba}{ {\bma{a}} }
\newcommand{\bB}{ {\bma{B}} }

\newcommand{\bd}{ {\bma{d}} }

\newcommand{\bH}{ {\bma{H}} }
\newcommand{\bh}{ {\bma{h}} }

\newcommand{\bX}{ {\bma{X}} }
\newcommand{\bx}{ {\bma{x}} }

\theoremstyle{definition}

\begin{document}

\title{Monte Carlo Sensitivity Analysis for unmeasured confounding in dynamic treatment regimes}


\pagestyle{empty}
\begin{center}
	\textbf{Monte Carlo Sensitivity Analysis for Unmeasured Confounding in Dynamic Treatment Regimes} \\
	\textbf{Eric J. Rose$^{1}$, Erica E. M. Moodie$^{1}$, 
		Susan Shortreed$^{2,3}$}
	\\
	$^1$Department of Epidemiology and Biostatistics, McGill University,
	Montreal, QC, H3A 1A2, Canada  \\
	$^2$Kaiser Permanente Washington Health Research Institute, Seattle, WA, 98101, U.S.A.\\
	$^3$Department of Biostatistics, University of Washington, Seattle, WA, 98195, U.S.A.
\end{center}

\begin{abstract} \noindent
Data-driven methods for personalizing treatment assignment have garnered much attention from clinicians and researchers. Dynamic treatment regimes formalize this through a sequence of decision rules that map individual patient characteristics to a recommended treatment. Observational studies are commonly used for estimating dynamic treatment regimes due to the potentially prohibitive costs of conducting sequential multiple assignment randomized trials. However, estimating a dynamic treatment regime from observational data can lead to bias in the estimated regime due to unmeasured confounding. Sensitivity analyses are useful for assessing how robust the conclusions of the study are to a potential unmeasured confounder. A Monte Carlo sensitivity analysis is a probabilistic approach that involves positing and sampling from distributions for the parameters governing the bias. We propose a method for performing a Monte Carlo sensitivity analysis of the bias due to unmeasured confounding in the estimation of dynamic treatment regimes. We demonstrate the performance of the proposed procedure with a simulation study and apply it to an observational study examining tailoring the use of antidepressants for reducing symptoms of depression using data from Kaiser Permanente Washington (KPWA). \\
\begin{footnotesize}
\end{footnotesize}
\end{abstract}

\pagebreak
\setcounter{page}{1}
\pagestyle{plain}

\section{Introduction}

Precision medicine focuses on data-driven methods for personalizing treatment assignment to improve health care. Dynamic treatment regimes (DTRs) operationalize clinical decision making through a sequence of functions that map individual patient characteristics to a recommended treatment \citep{Chakraborty_Moodie, dtr_book}. An optimal treatment regime is one that maximizes the mean of some desirable measure of clinical outcome when applied to select treatment for all patients in the population of interest \citep{murphy2003optimal, robins2004optimal}. Optimal treatment regimes have been estimated to improve care for many different medical conditions including HIV \citep{van2005history, cain2010start}, bipolar disorder \citep{wu2015will, listBasedRegimes}, diabetes \citep{ertefaie_2018, luckett_2018}, and cancer \citep{thall2000evaluating, thallBayesQ}.

Data for estimating an optimal DTR is ideally collected through the use of a sequential multiple assignment randomized trial (SMART) \citep{lavori2000design, lavori2004dynamic, murphy2005experimental}. However, longitudinal observational studies are frequently used due to the availability of electronic health records and the potentially prohibitive cost of conducting a SMART. These analyses rely on assuming that there is no unmeasured confounding; this assumption is not verifiable from the observed data. Not adjusting for all of the confounding variables can result in a biased estimate of the optimal treatment regime. 

Sensitivity analysis has been used as a way to assess how the estimated effect would change if there was an unmeasured confounder. There has been much work on conducting sensitivity analysis for unmeasured confounding when estimating average causal effects from observational data beginning with \cite{Cornfield_1959}, in which they studied whether the causal link found between smoking and lung cancer could be due to unmeasured confounding. A wide array of different approaches have been proposed since. \cite{Rosenbaum_1983} proposed a method for examining how sensitive the conclusions of a study with a binary outcome are to a binary unmeasured confounder by assessing what odds ratios between the unmeasured confounder and each of the treatment and the outcome would cause the effect to be no longer significant. \cite{Lin_1998} proposed positing a regression model for the outcome containing the effect of the unmeasured confounder and probability distributions for the unmeasured confounder in each treatment group. Under certain assumptions, this allows for simple expressions of the bias in the treatment effect which can then be used to calculate treatment effects for a wide range of different effect sizes of the unmeasured confounder on the outcome and exposure.

Probabilistic sensitivity analysis is an alternative to formulaic approaches that works by positing probability distributions for the bias parameters and averaging over the distributions to obtain bias-adjusted estimates. There are two different approaches to implementing probabilistic sensitivity analysis, Bayesian sensitivity analysis and Monte Carlo sensitivity analysis \citep{McCandless_2017}. Bayesian sensitivity analysis \citep{McCandless_2007} works by positing prior distributions for the bias parameters and uses Bayes theorem to generate a posterior distribution for the causal effect that accounts for the uncertainty due to the unmeasured confounder. In Monte Carlo sensitivity analysis \citep{Phillips_2003, Steenland_2004, Greenland_2005} we posit prior distributions for the bias parameters in the same way and then draw values from the prior distributions and calculate a bias-adjusted estimate for the parameter of interest for each Monte Carlo repetition. This allows for a straightforward implementation that corrects bias while incorporating uncertainty due to the unmeasured confounding within a frequentist framework.

There has been some work on estimating DTRs in the presence of unmeasured confounding. This has mainly focused on using instrumental variables to estimate an optimal treatment regime \citep{Chen_2021, Liao_2021, Qiu_2021, Cui_2021}. An instrumental variable is a pre-treatment variable that is correlated with treatment, independent of all unmeasured confounders, and has no direct effect on the outcome of interest. In some applications, there is not an obvious choice for an instrumental variable based on domain knowledge and it is impossible to empirically check whether a variable satisfies the assumptions for being an instrumental variable \citep{hernan_2006}. \cite{Zhang_2021} proposed a method for ranking individualized treatment rules when there is unmeasured confounding by creating a partial order under the framework proposed by \cite{Rosenbaum_1987}. \cite{Kallus_2019} examined estimating a treatment regime that maximizes the value for the worst case scenario of an uncertainty set that quantifies the degree of confounding that is unmeasured.

We propose using a Monte Carlo sensitivity analysis for the estimation of DTRs. Section \ref{s:setup} details setup and notation.  In Section \ref{s:mcsa}, we provide an overview of our proposed sensitivity analysis procedure. In Section \ref{s: sims}, we examine the performance of our proposed method using a simulation study. Section \ref{s:ehrs} demonstrates applying the proposed procedure to electronic health records (EHRs) when estimating treatment regimes to reduce symptoms of depression using data from Kaiser Permanente Washington (KPWA). A discussion of the proposed procedure is contained in Section \ref{s:discussion}.

\section{Setup and notation}
\label{s:setup}
Suppose we have data from an observational study in which patients are treated throughout the course of $K$ stages. We collect data of the form $\mathcal{D}_n = \{(\bX_{1,i}, A_{1,i}, \dots, \allowbreak \bX_{K,i}, A_{K,i}, Y_i) \}_{i=1}^{n}$ which consists of $n$ $i.i.d.$ replicates of $(\bX_{1}, A_{1}, \dots, \allowbreak \bX_{K}, A_{K}, Y) $ such that $\bX_1 \in \mathbb{R}^{p_1}$ denotes baseline patient characteristics, $A_k \in \{0,1\}$ denotes the treatment assigned at stage $k$ for $k=1,\dots,K$, $\bX_k \in \mathbb{R}^{p_k}$ for $k = 2,\dots,K$ denotes patient information recorded during the course of the $(k-1)^{\mathrm{st}}$ treatment, and $Y \in \mathbb{R}$ denotes the patient outcome coded such that higher values are better. We let $U$ denote an unmeasured time-fixed confounding variable that can be continuous or binary. This unmeasured confounder is measured at baseline and does not change over time, but influences treatment choices at any of the $K$ stages as well as the outcome $Y$. Let $\bH_1 = \bX_1$ and $\bH_k = (\bX_1, A_1, \dots, A_{k-1}, \bX_k)$ for $k = 2, \dots, K$ denote the history of a patient such that $\bH_k$ is all of the information available to a clinical decision maker at stage $k$. Let $\bd = (d_1, \dots, d_K)$ denote a treatment regime such that $d_k: \mathrm{dom}~\bH_k \rightarrow \mathrm{dom}~A_k$ for $k=1,\dots,K$ is a function that maps a patient's history to a recommended treatment, where $\mathrm{dom}$ indicates the domain or set of possible treatments. Therefore treatment regime $\bd$ would recommend treatment $d_k(\bh_k)$ at stage $k$ for a patient with history $\bH_k = \bh_k$. 

To allow us to reference components of the history and regimes, we will use overbar notation to indicate up to stage $k$ such that $\bar{\bx}_k = (\bx_1, \dots, \bx_k)$, $\bar{\ba}_k = (a_1, \dots, a_k)$, and $\bar{\bd}_k = (d_1, \dots, d_k)$. We will suppress the subscript when denoting the entire sequence, e.g. $\bar{\bx} = \bar{\bx}_K$. Similarly, we will use an underbar to denote treatments, covariates, and regimes after and including stage $k$ such that $\underline{\bx}_k = (\bx_k, \dots, \bx_K)$ and similarly for $\underline{\ba}_k$ and $\underline{\bd}_k$.

We will use the potential outcomes framework to define an optimal treatment regime \citep{rubin_1978}. Let $Y^*(\bar{\ba})$ denote the potential outcome under the treatment sequence $\bar{\bA} = \bar{\ba}$ and $\bH_k^*(\bar{\ba}_{k-1})$ denote the potential history at stage $k$ under treatments $\bar{\bA}_{k-1} = \bar{\ba}_{k-1}$. Denote the set of all potential outcomes as $\bma{\mathcal{O}}^* = \{ \bH_2^*(a_{1}), \bH_3^*(\bar{\ba}_{2}), \dots, \bH_K^*(\bar{\ba}_{K-1}), Y^*(\bar{\ba}) : \bar{\ba} \in \{0,1\}^K \}$. The potential outcome of a treatment regime, $\bd$, is then defined as
$$
Y^*(\bd) = \sum_{\bar{\ba} \in \{0,1\}^K} Y^*(\bar{\ba}) \mathbb{I}\{d_1(\bh_1) = a_1\} \prod_{k=2}^{K} \mathbb{I}\left[d_k\{\bH_k^*(a_{k-1})\} = a_k \right]
$$
where $\mathbb{I}$ is the indicator function. Define the value of a treatment regime by $V(\bd) = \E\{ Y^*(\bd) \}$. An optimal regime, $\bd^{opt}$, is defined as a regime that satisfies
$ V(\bd^{opt}) \geq V(\bd) $ for all regimes $\bd$.

When estimating DTRs, it is standard to make the following causal assumptions \citep{robins2004optimal}: (i) the stable unit treatment value assumption (SUTVA), $Y = Y^*(\bar{\bA})$ and $\bH_k = \bH_k^*(\bar{\bA}_{k-1})$ for $k=2,\dots,K$; (ii) positivity, $\Pr(A_k = a_k | \bH_k = \bh_k) > 0$ with probability one for each $a_k \in \{0,1\}$ for $k=1,\dots,K$; and (iii) sequential ignorability, $ \bma{\mathcal{O}}^* \perp A_k | \bH_k $ for $k=1,\dots,K$. We will assume SUTVA and positivity holds, but sequential ignorability does not hold because we have an unmeasured confounder, $U$. This is a key assumption and if it is violated, standard methods for estimating DTRs will lead to biased estimates of the optimal treatment regime. In addition, sequential ignorability is unverifiable from data and requires domain expertise to assess whether it is a reasonable assumption. We will assume that $U$ is the only unmeasured confounder and $\bma{\mathcal{O}}^* \perp A_k | U, \bH_k $ for $k=1,\dots,K$. We will also assume that the unmeasured confounder has an additive effect on the outcome $Y$ and is conditionally independent of later stage covariates so we have that $\bX_k \perp U | \bH_{k-1}, A_{k-1}$ for $k = 2,\dots, K$.

\subsection{Dynamic weighted ordinary least squares (dWOLS)}

Dynamic weighted ordinary least squares (dWOLS) is a regression-based method for estimating DTRs \citep{Wallace_2015}. This method estimates an optimal treatment regime by estimating a contrast function that represents the difference in expected outcome of receiving a given treatment at stage $k$ when compared to the reference treatment, which we take to be treatment $A_k = 0$, assuming treatment is assigned optimally after stage $k$.
This contrast function is referred to as the optimal blip-to-zero function, $\gamma(\bh_k, a_k)$, and is defined as 
\begin{align*}
	&\gamma_K(\bh_K, a_K) = \E \{Y^*(\bar{\ba}_{K-1}, a_K)- Y^*(\bar{\ba}_{K-1}, 0) | \bH_K = \bh_K \}, \\
	&\gamma_k(\bh_k, a_k) = \E \left\{  Y^*\left( \bar{\ba}_{k-1}, a_k, \underline{\bd}_{k+1}^{opt} \right)  - Y^* \left(\bar{\ba}_{k-1}, 0, \underline{\bd}_{k+1}^{opt} \right) \vert  \bH_k = \bh_k \right\} ~\mathrm{for}~ k = 2,\dots,K-1, \\
	&\gamma_1(\bh_1, a_1) = \E \left\{  Y^*\left( a_1, \underline{\bd}_{2}^{opt} \right) - Y^* \left( 0, \underline{\bd}_{2}^{opt} \right) \vert  \bH_1 = \bh_1 \right\}.
\end{align*}
The optimal blip-to-zero function characterizes the optimal treatment regime. The optimal treatment rule at stage $k$ is given by recommend treatment $A_k = 1$ if $\gamma_k(\bh_k, 1) > 0$, otherwise recommend $A_k = 0$.
In general, when considering more than two treatment options, the optimal treatment regime at stage $k$ is given by 
$
d_k^{\mathrm{opt}}(\bh_k) = \arg\max_{a_k} \gamma_k(\bh_k, a_k).
$ 
dWOLS estimates an optimal treatment regime by positing models for the blip functions and estimating their parameters through a series of weighted ordinary least squares regressions. Let $\gamma_k(\bh_k, a_k; \psi_k)$ denote our posited model for the blip function. We will assume that the blip-to-zero function is correctly specified so we have that $\gamma_k(\bh_k, a_k; \psi_k^*) = \gamma_k(\bh_k, a_k) $ for all $k$. The estimated optimal treatment regime is then given by $\hat{\bd}^{\mathrm{opt}} = (\hat{d}_1^{\mathrm{opt}}, \dots, \hat{d}_K^{\mathrm{opt}})$ such that  
$ \hat{d}_k^{\mathrm{opt}}(\bh_k) = \arg\max_{a_k} \gamma_k(\bh_k, a_k; \hat{\psi}_k)$. 

Denote the treatment-free outcome at stage $k$ by $G_k(\underline{\psi}_k)$ which represents a patient's observed outcome adjusted by the expected difference in a patient's outcome had they received treatment 0 at stage $k$ and then treated optimally via the blip models with parameters $\underline{\psi}_k$ after stage $k$. 
Therefore $G_k(\underline{\psi}_k) = Y - \gamma_k(\bh_k, a_k;\psi_k) + \sum_{j=k+1}^{K} \left[ \gamma_j\{\bh_j, d_j^\mathrm{opt}(\bh_j);\psi_j\} - \gamma_j(\bh_j, a_j;\psi_j) \right]$. We posit a model for the treatment-free outcome which we denote by $g_k(\bh_k, a_k; \beta_k)$. 
We assume linear models for the treatment-free and blip models so a model for the outcome at stage $K$ is given by
\begin{align*}
	\E(Y|\bH_K = \bh_K, A_K = a_K; \beta_K, \psi_K) & = g_K(\bh_K, a_K; \beta_K) + \gamma_K(\bh_K, a_K; \psi_K) \\
	& = \bh_{K,\beta}^T\beta_K + a_K\bh_{K,\psi}^T\psi_K
\end{align*}
where $\bh_{K,\beta}$ and $\bh_{K,\psi}$ are components of $\bh_K$ with a leading one for the intercept and $\bh_{K,\psi} \subset \bh_{K,\beta}$. 
We also model the propensity score, $\pi_k(\bh_k) = \E(A_k|\bH_k = \bh_k)$, which we denote by $\pi_k(\bh_k; \xi_k)$  for $k=1,\dots,K$. 
We estimate the blip model parameters using weighted ordinary least squares using weights $w_k(a_k, \bh_k; \xi_k)$ that satisfy $\pi_k(\bh_k; \xi_k) w_k(1, \bh_k; \xi_k) = \{1 - \pi_k(\bh_k; \xi_k)\}w_k(0, \bh_k; \xi_k)$. Examples of weights that satisfy this equality are given by the inverse probability of treatment weights (IPTW), $a_k \{\pi(\bh_k; \xi_k)\}^{-1} + (1 - a_k)\{1 - \pi(\bh_k; \xi_k)\}^{-1}
$ 
and overlap weights,  $|a_k - \pi(\bh_k; \xi_k)|$. First, we estimate the blip parameters for the final stage $K$ using our posited model for the outcome. Then form the pseudo-outcome which we define as
$$ \tilde{Y}_k = Y + \sum_{j=k+1}^{K} \left[ \gamma_j\{\bh_j, \hat{d}_j^\mathrm{opt}(\bh_j);\hat{\psi}_j\} - \gamma_j(\bh_j, a_j;\hat{\psi_j}) \right]. $$
At each stage starting with $k = K-1$ and moving backwards, we model the pseudo-outcome by
$$
\E(\tilde{Y}_k|\bH_k = \bh_k, A_k = a_k; \beta_k, \psi_k) = g_k(\bh_k, a_k; \beta_k) + \gamma_k(\bh_k, a_k; \psi_k)
$$
and fit the model using weighted least squares. The resulting estimate of $\psi$ is consistent as long as the blip model is correctly specified and at least one of the treatment free and propensity score models is correctly specified. When we have an unmeasured confounder, $U$, both the treatment free and propensity score model will be misspecified leading to bias in the estimator of $\psi$.

\section{Monte Carlo Sensitivity Analysis}
\label{s:mcsa}

Monte Carlo sensitivity analysis is a probabilistic approach to sensitivity analysis in which probability distributions are posited for the parameters governing the bias due to the unmeasured confounder. The posited probability distributions are repeatedly sampled from and used to calculate a series of bias corrected effect estimates \citep{Greenland_2005, McCandless_2017}. In what follows, we detail a Monte Carlo approach that seeks to posit models (and associated parameters) to describe the confounding due to the unmeasured variable $U$ through its relationship to the outcome, the treatment, and other covariates, and to use that to impute the unmeasured confounder and hence assess and correct bias. Ideally, an external data set containing the unmeasured confounder should be used to help posit these models. If an external data set is unavailable, these can be posited using subject matter knowledge, but wider distributions should be used for the bias parameters to capture the additional uncertainty.

The primary objective of conducting sensitivity analysis in this context is to quantify the uncertainty in the estimated treatment regime when there is bias due to unmeasured confounding. If we generated confidence intervals for the bias by taking the percentiles of our adjusted results across the Monte Carlo samples, these interval estimates would only capture the variability due to the uncertainty in our bias parameters and would not account for the uncertainty in our estimate of the parameters of interest. Therefore, we take a bootstrap sample for each Monte Carlo repetition and confidence intervals are formed by taking percentiles of the bias adjusted values of $\psi$. To account for non-regularity in estimators resulting from the non-smoothness in the pseudo-outcome when estimating a multi-stage DTR, we will adapt the $m$-out-of-$n$ bootstrap \citep{Bibhas_2013}.

Let $\bB_k = (\bH_{k,\beta}^T, A_k\bH_{k,\psi}^T)^T$ and $\hat{w}_k = w_k(A_k, \bH_k; \hat{\xi}_k)$ for $k=1,\dots,K$. If the unmeasured confounder $U$ is not included in the regression, then the estimated coefficients are given by 
\begin{align*}
	\begin{pmatrix}
		\hat{\beta}_K \\ \hat{\psi}_K 
	\end{pmatrix} = \left\{ \mathbb{P}_n \left( \hat{w}_K\bB_K\bB_K^T \right) \right\}^{-1} \mathbb{P}_n \left( \hat{w}_K\bB_KY \right) 
\end{align*}
such that $\mathbb{P}_n$ denotes the empirical expectation. Recall we assume the unmeasured confounder, $U$, has an additive effect on the outcome and let $\beta_u$ denote the coefficient for the unmeasured confounder. In Section A of the supplementary material, we show the bias is given by
\begin{align}
	\label{bias_formula}
	\mathrm{bias}(\hat{\psi}_K) & = \E(\hat{\psi}_K) - \psi_K \nonumber \\
	& = \bigg[-  \E \left( \hat{w}_K A_K \bH_{K,\psi}\bH_{K,\beta}^T \right) \left\{\E \left( \hat{w}_K \bH_{K,\beta} \bH_{K,\beta}^T \right) \right\}^{-1} 
	\E \left( \hat{w}_K A_K \bH_{K,\beta} \bH_{K,\psi}^T \right)	 
	\nonumber \\ & \qquad \quad
	+  \E \left( \hat{w}_K A_K \bH_{K,\psi} \bH_{K,\psi}^T \right)																					
	\bigg]^{-1} 
	\bigg[ \E \left( \hat{w}_K A_K \bH_{K,\psi}\beta_u U \right) 
	\\ & \qquad \quad
	-\E \left( \hat{w}_K A_K \bH_{K,\psi} \bH_{K,\beta}^T \right) \left\{ \E \left( \hat{w}_K \bH_{K,\beta} \bH_{K,\beta}^T \right) \right\}^{-1} 
	\E \left( \hat{w}_K \bH_{K,\beta}\beta_u U \right) 
	\bigg]. \nonumber
\end{align}
We can estimate the bias in finite samples by replacing the expectations with empirical expectations and imputing an estimate of $\E(U|\bH_K, A_K)$ for $U$ which we will denote by $\widehat{\mathrm{bias}}(\hat{\psi}_K)$. Note that this is then a function of the data and the parameters of the unmeasured confounder models. 

If we proceeded with dWOLS, the bias in the blip parameters would bias the pseudo-outcome given 
by $\tilde{Y}_{K-1} =  Y + \gamma_{K}\{\bh_{K}, \hat{d}_{K}^\mathrm{opt}(\bh_{K});\hat{\psi}_{K}\} - \gamma_{K}(\bh_{K}, a_{K};\hat{\psi}_{K}) $. Additionally, estimating the blip model at stage $K-1$ without accounting for the unmeasured confounder would introduce additional bias that further compounds as we move backwards through the stages. We can use the bias corrected estimate of $\psi$ to calculate the pseudo-outcome and, if we assume that $\bX_k \perp U | \bH_{k-1}, A_{k-1}$ for $k=2,\dots,K$, we have that equation \ref{bias_formula} for $k=K$ holds for $k = 1,\dots, K$. 

We perform Monte Carlo sensitivity analysis by following algorithms \ref{algo_mcsa} and \ref{algo_ci}.
Algorithm \ref{algo_mcsa} displays the steps for generating bias adjusted estimates for the parameters indexing the optimal regime while continuing with algorithm \ref{algo_ci} constructs a confidence interval for $\psi$ that accounts for uncertainty in the unmeasured confounder and random sampling variability.

\begin{algorithm}
	\caption{Monte Carlo sensitivity analysis}
	\label{algo_mcsa}
	\begin{algorithmic}[1]
		\State Posit a model for $\E(U|\bH_K = \bh_K, A_K = a_K)$ given by $\E(U|\bH_K = \bh_K, A_K = a_k; \zeta)$
		\State Posit distributions for $\zeta, \beta_u$
		\For{$b \in \{1, \dots, B\}$ Monte Carlo/bootstrap repetitions}
		\State Draw a bootstrap sample from $\mathcal{D}_n$
		\State Sample $\hat{\zeta}^{(b)}, \hat{\beta}_{u}^{(b)}$ from the distributions posited in step 2
		\State Impute $\hat{U}^{(b)}$ from $\E\left(U|\bH_K = \bh_K, A_K = a_K; \hat{\zeta}^{(b)}\right)$
		\State Estimate $\hat{\xi}_K^{(b)}$ with the propensity score model and calculate weights $\hat{w}_K^{(b)}$
		\State \parbox[t]{\dimexpr\linewidth-\algorithmicindent}{Estimate $\hat{\psi}_K^{(b)}$ using weighted least squares, i.e. perform a dWOLS estimation using the bootstrap sample \strut}
		\State \parbox[t]{\dimexpr\linewidth-\algorithmicindent}{Calculate bias in $\hat{\psi}_K^{(b)}$ with equation \ref{bias_formula} using $\hat{U}^{(b)}$ and $\hat{\beta}_{u}^{(b)}$ as estimates for $U$ and $\beta_u$ \strut}
		\State Calculate bias adjusted estimate $\hat{\psi}_K^{\mathrm{adj},(b)} = \hat{\psi}_K^{(b)} - 
		\widehat{\mathrm{bias}}(\hat{\psi}_K^{(b)}) $ 
		\For{$k = K-1, \dots, 1$}
		\State Calculate pseudo-outcome 
		$$ \tilde{Y}_k^{(b)} = Y + \sum_{j=k+1}^{K} \left[ \gamma_j\{\bh_j, \hat{d}_j^\mathrm{opt}(\bh_j);\hat{\psi}_j^{\mathrm{adj},(b)}\} - \gamma_j(\bh_j, a_j;\hat{\psi}_j^{\mathrm{adj},(b)}) \right] $$
		\State Estimate $\hat{\xi}_k^{(b)}$ with the propensity score model and calculate weights $\hat{w}_k^{(b)}$
		\State Estimate $\hat{\psi}_k^{(b)}$ using weighted least squares
		\State Calculate bias in $\hat{\psi}_k^{(b)}$ with equation \ref{bias_formula} using $\hat{U}^{(b)}$ and $\hat{\beta}_{u}^{(b)}$ as estimates for $U$, $\beta_u$
		\State Calculate bias adjusted estimate $\hat{\psi}_k^{\mathrm{adj},(b)} = \hat{\psi}_k^{(b)} - 
		\widehat{\mathrm{bias}}(\hat{\psi}_k^{(b)}) $	
		\EndFor
		\EndFor
		\State Calculate bias adjusted estimate $\hat{\psi}_k^{\mathrm{adj}} = B^{-1}\sum_{b=1}^{B} \hat{\psi}_k^{\mathrm{adj},(b)}$ for $k=1,\dots,K$
	\end{algorithmic}	
\end{algorithm}

\begin{algorithm}
	\caption{Monte Carlo sensitivity analysis: confidence intervals}
	\label{algo_ci}
	\begin{algorithmic}[1]
		\State Calculate $\hat{p}_K = \mathbb{P}_n \mathbb{I}\{n(\bH_{K,\psi}^T\hat{\psi}_K^{\mathrm{adj}})^2 \leq \bH_{K,\psi}^T\hat{\Sigma}_{\psi_K}\bH_{K,\psi}\chi^2_{1, 1-\nu} \}$ where $\hat{\Sigma}_{\psi_K}$ denotes an estimate of $n\mathrm{Cov}(\hat{\psi}_K, \hat{\psi}_K)$
		\State Calculate resample size $\hat{m}_{K} = n^{\frac{1 + \kappa\{1-\hat{p}_{K}\}}{1 + \kappa}}$ 
		\For{$k= K-1, \dots, 1$}
		\State Draw $B$ bootstrap samples of size $\hat{m}_{k}$
		\State Estimate $\hat{\psi}_{k, \hat{m}_k}^{\mathrm{adj}, (b)}$ for $b=1,\dots,B$
		\State For each given $\bH_{k,\psi} = \bh_{k,\psi}$, construct a confidence interval for $\bh_{k,\psi}^T\psi_{k}^{\mathrm{adj}}$
		\State Calculate $\hat{p}_k$ by the proportion of confidence intervals for $\bh_{k,\psi}^T\psi_{k}^{\mathrm{adj}}$ that contain zero
		\State Calculate the resample size $\hat{m}_{k} = n^{\frac{1 + \kappa\{1-\hat{p}_{k}\}}{1 + \kappa}}$
		\EndFor
		\State Calculate $\hat{p} = \max_k \hat{p}_k $ and $\hat{m} = n^{\frac{1 + \kappa\{1-\hat{p}\}}{1 + \kappa}}$
		\State Draw $B$ bootstrap sample of size $\hat{m}$
		\State  Estimate $\hat{\psi}_{k, \hat{m}_k}^{\mathrm{adj}, (b)}$ for $k=K,\dots,1$, $b=1,\dots,B$
		\State Calculate $\vartheta_k/2 \times 100$ and $(1- \vartheta_k/2) \times 100$ percentiles of $\sqrt{\hat{m}_K}(\hat{\psi}_{k, \hat{m}_k}^{\mathrm{adj},(b)} - \hat{\psi}_k^{\mathrm{adj}})$ which we denote by $\hat{l}_k$ and $\hat{u}_k$ respectively for $k=1,\dots,K$
		\State Calculate a confidence interval for $\psi_k$ by $(\hat{\psi}_{k}^{\mathrm{adj}} - \hat{u}_k/\sqrt{\hat{m}}, \hat{\psi}_{k}^{\mathrm{adj}} + \hat{l}_k/\sqrt{\hat{m}})$ for $k=1,\dots,K$
	\end{algorithmic}	
\end{algorithm}

\section{Simulation Experiments}
\label{s: sims}

We conducted a series of simulation experiments to evaluate the effectiveness of the proposed method. We applied the proposed method to two different data generating models, the first being a one-stage study and the second a two-stage study. Additional simulations are contained in Section B of the supplemental material in which the bias does not significantly effect the performance of the estimated regime. The data generating model for the one-stage study was given by:
\begin{equation*}
	\begin{array}{ll}
		U \sim N(0, \sigma_u^2) \\
		X_1 = \phi_{10} + \phi_{11}U + \epsilon_{x_1}  & ~~~~~  \epsilon_{x_1} \sim N(0, \sigma_{x_1}^2)  \\
		X_2 = \phi_{20} + \phi_{21}U + \epsilon_{x_2}  & ~~~~~  \epsilon_{x_2} \sim N(0, \sigma_{x_2}^2)  \\
		\Pr(A = 1|\bX = \bx, U = u) = \left[1 + \exp \left\{-(\alpha_{0} + \alpha_{1}x_1 + \alpha_{2}x_2 + \alpha_{3}u)\right\}\right]^{-1}  \\ 
		Y = \beta_{0} + \beta_{1}X_1 + \beta_{2}X_2 + \beta_{u}U + A(\psi_{0} + \psi_{1}X_1 + \psi_{2}X_2) + \epsilon_y  & ~~~~~ \epsilon_y \sim N(0, \sigma_y^2).  \\
	\end{array}
\end{equation*}
We conducted 1000 repetitions for the simulation study. We let the sample size be given by $n=1000$ and conducted $B = 500$ Monte Carlo repetitions for the sensitivity analysis.
The parameter values were given by:
\begin{equation*}
	\begin{array}{ll}
		\psi = (\psi_0, \psi_1, \psi_2) =  (-1, 0.5, 0.5), & \alpha = (\alpha_0, \alpha_1, \alpha_2, \alpha_{3})  = (0, 1, 1, 2),  \\
		\phi_1 = (\phi_{10}, \phi_{11}) = (0, 1), & \beta = (\beta_0, \beta_1, \beta_2, \beta_u) = (1, 1, 1, 2),   \\
		\phi_2 = (\phi_{20}, \phi_{21}) = (0, -1), &  \sigma_u^2 = \sigma_{x_1}^2 = \sigma_{x_2}^2 = \sigma_y^2 = 1.  \\
	\end{array}
\end{equation*}

We posited four different sets of distributions for the parameters in the unmeasured confounder models. This allowed us to asses how sensitive our proposed method is to bias in the distributions for these parameters. The four different simulation scenarios were given by: (i) narrow normal, centered properly; (ii) wide normal, centered properly; (iii)  narrow normal, off-center; (iv) wide normal, off-center. For scenario (i), we posited models given by $\hat{\beta}_{u}^{(m)} \sim N(\beta_{u}, 0.1)$ and $\hat{\zeta}_{j}^{(m)} \sim N(\zeta_{j}^*, 0.1) \mathrm{~for~} j = 0,1,2$. For the wide distribution settings, the variance was increased to 0.5 and for the off-center scenarios the distribution was centered at the true mean plus 0.1.

\begin{figure}[h]
	\centering
	\includegraphics[width=0.95\linewidth]{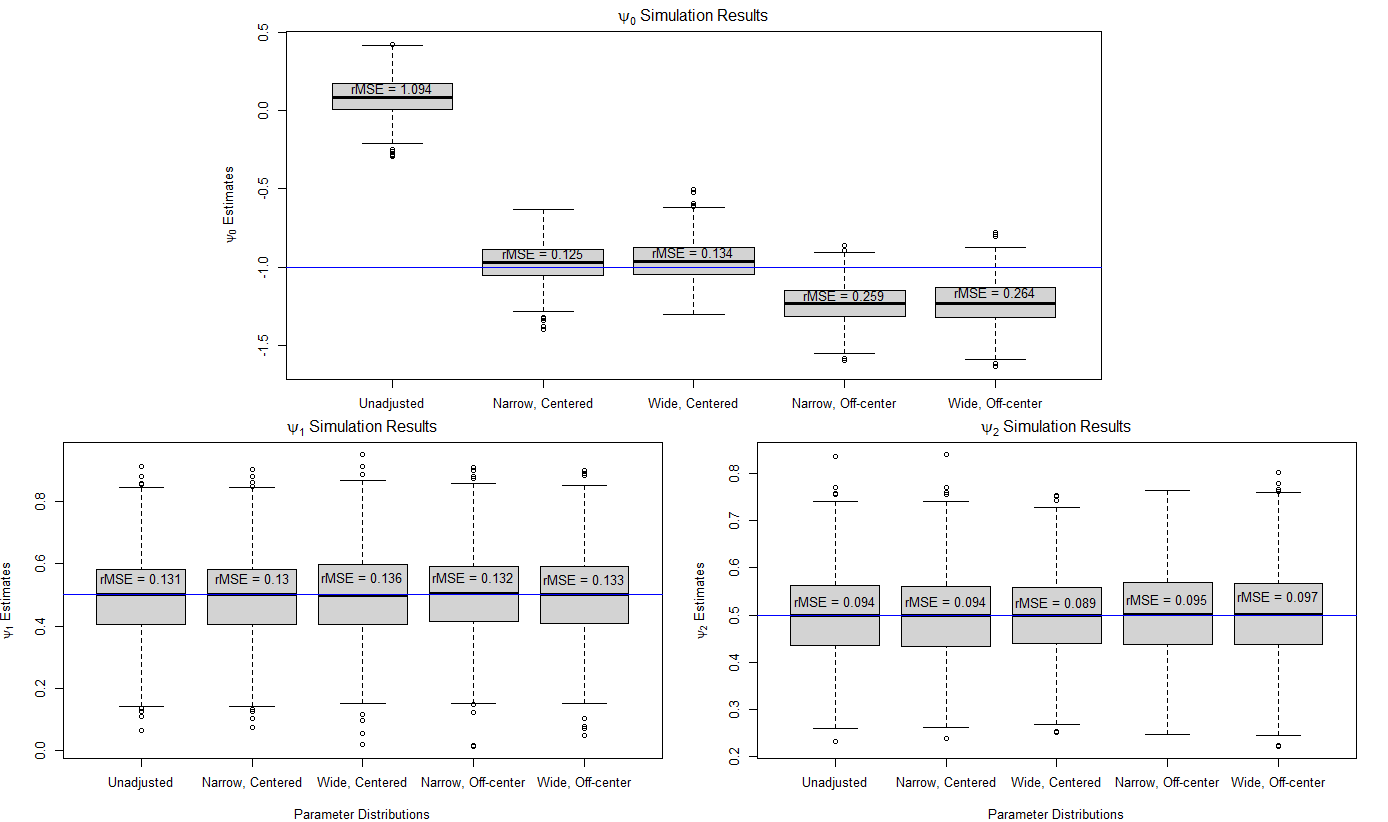}
	\caption{Boxplots of the point estimates for $\psi$ under an unadjusted model and when using Monte Carlo sensitivity analysis to adjust for bias due to unmeasured confounding for the 1-stage data generating model.}
	\label{fig:1stage_psi}
\end{figure}

Figure \ref{fig:1stage_psi} shows boxplots of the point estimates of $\psi_0$, $\psi_1$, and $\psi_2$ across 1000 repetitions when not adjusting and adjusting for the unmeasured confounder using Monte Carlo sensitivity analysis with different distributions for the bias parameters. We can see that when we did not adjust for bias due to the unmeasured confounder, the estimate of $\psi_0$ is biased with a root mean squared error (rMSE) of 1.094. When the parameter distributions were centered on the true value, the adjusted estimate of $\psi_0$ was unbiased. There was some bias in the adjusted values of $\psi_0$ when the distribution is off-center as expected. We can see that for the narrow distribution the rMSE increased from 0.125 to 0.259 and for the wide distribution rMSE increased from 0.134 to 0.264. The unmeasured confounder did not cause bias in the estimation of $\psi_1$ and $\psi_2$ so the adjusted estimates using Monte Carlo sensitivity analysis were very similar to the unadjusted results.

Bias in the blip model parameters that index the optimal DTR can cause the estimated regime to differ from the true optimal regime. For each repetition of the simulation study, we simulated 10000 new patients and assessed what proportion of the patients' treatment would match the recommended treatment of the true optimal treatment regime when following each one of the adjusted and unadjusted estimates of the optimal regime. Table \ref{propOpt_1stage} displays the proportion of patients who would receive the treatment recommend by the true optimal regime across all 1000 repetitions of the simulation study. We can see that the bias in $\psi_0$ causes the treatment recommended by the estimated regime to match the optimal regime 52.8\% of the time. When adjusting for the unmeasured confounder this increased to 95.6\% in the ideal scenario of a narrow, centered parameter distribution and to 95.8\% when we used a wide, off-centered parameter distributions. 

\begin{table}[h]
	\caption{
		Proportion of patients whose recommended treatment when following each of the estimated regime matches the recommendation of the true optimal regime for the 1-stage data generating model.
	}
	\label{propOpt_1stage}
	\begin{center}
		\begin{tabular}{|c c|} \hline
			Parameter Distr. & Proportion Optimal \\ \hline
			Unadjusted & 0.528 \\ 
			Narrow, Centered & 0.956 \\ 
			Wide, Centered & 0.953 \\ 
			Narrow, Off-center & 0.959 \\ 
			Wide, Off-center & 0.958 \\   \hline
		\end{tabular}
	\end{center}
\end{table}

In practice, if we do not have external data to help posit the mean model for the unmeasured confounder and distributions for the parameters, we will have less belief that the bias adjusted treatment regime is close to the true optimal regime. The proposed method should be used to examine confidence intervals to assess whether the unmeasured confounder could be introducing bias to our estimated regimes. Table \ref{cover_1stage} displays the coverage and width of 95\% confidence intervals generated using Monte Carlo sensitivity analysis and from the unadjusted analysis that assumes no unmeasured confounding. For the unadjusted analysis, the empirical converage of the confidence interval for $\psi_0$ was 0\%. Even for $\psi_1$ and $\psi_2$, which did not appear to have any bias in the estimation, the coverage was still below the nominal rate of 95\%. For all of the different posited parameter distributions, Monte Carlo sensitivity analysis produced confidence intervals that attained or were close to the nominal coverage probability. The width of the confidence intervals for $\psi_0$ increased significantly when conducting sensitivity analysis, indicating that this parameter is sensitive to the unmeasured confounding. Note that the width of the confidence intervals increased when the variability in the parameter distributions was increased.

\begin{table}[h]
	\caption{
		Coverage (Cvr.) and average width (Wth.) of the 95\% confidence intervals for $\psi$ for the unadjusted analysis and sensitivity analysis under each of the posited parameter distributions for the 1-stage data generating model. * indicates coverages that are significantly different than 95\%.
	}
	\label{cover_1stage}
	\begin{center}
		\begin{tabular}{|c c c c|} \hline
			Parameter Distr. & Cvr. (Wth.) $\psi_0$ & Cvr. (Wth.)  $\psi_1$ & Cvr. (Wth.) $\psi_2$ \\ \hline
			Unadjusted & 0.000* (0.359) & 0.897* (0.426) & 0.887* (0.300) \\ 
			Narrow, Centered & 1.000* (2.628) & 0.946 (0.520) & 0.947 (0.365) \\ 
			Wide, Centered & 1.000* (6.212) & 0.954 (0.520) & 0.959 (0.365) \\ 
			Narrow, Off-center & 1.000* (2.764) & 0.950 (0.519) & 0.951 (0.366) \\ 
			Wide, Off-center & 1.000* (6.495)  & 0.955 (0.519) & 0.936* (0.367) \\  \hline
		\end{tabular}
	\end{center}
\end{table}

We conducted a similar simulation study with a 2-stage study to assess the effectiveness of Monte Carlo sensitivity analysis for the estimation of multi-stage DTRs. We let the data generating model be given by:
\begin{equation*}
	\begin{array}{ll}
		U \sim N(0, \sigma_u^2), \\
		X_{11} = \phi_{10} + \phi_{11}U + \epsilon_{x_{11}},  &  \epsilon_{x_{11}} \sim N(0, \sigma_{x_{11}}^2),  \\
		X_{12} = \phi_{20} + \phi_{21}U + \epsilon_{x_{12}},  &  \epsilon_{x_{12}} \sim N(0, \sigma_{x_{12}}^2),  \\
		X_2 = \varpi_{0} + \varpi_{1}X_{11} + \varpi_{2}X_{12} + \epsilon_{x_2},  & \epsilon_{x_2} \sim N(0, \sigma_{x_2}^2),  \\
		\multicolumn{2}{l}{\Pr(A_1 = 1|\bX_1 = \bx_1, U = u) = \left[1 + \exp\left\{-(\alpha_{10} + \alpha_{11}x_{11} + \alpha_{12}x_{12} + \alpha_{13}u)\right\} \right]^{-1},} \\ 
		\multicolumn{2}{l}{\Pr(A_2 = 1|\bH_2 = \bh_2, U = u) = \left[1 + \exp\left\{-(\alpha_{20} + \alpha_{21}x_{11} + \alpha_{22}x_{12} + \alpha_{23}a_1 + \alpha_{24}x_2 + \alpha_{25}u)\right\} \right]^{-1},}
		 \\ 
		Y = \beta_{20} + \beta_{21}X_{11} + \beta_{22}X_{12} + \beta_{23}A_1 + \beta_{24}A_1 X_{11} + \beta_{25}A_1 X_{12} \\ \qquad + \beta_{26} X_2  + \beta_{u} U  + A_2(\psi_{20} + \psi_{21}X_{11} + \psi_{22}X_{12} + \psi_{23} X_2 ) + \epsilon_y,  & 
		\epsilon_y \sim N(0, \sigma_y^2).  \\
	\end{array}
\end{equation*}
As before, we conducted 1000 repetitions for the simulation study, set the sample size to $n=1000$, and conducted the sensitivity analysis using $B = 500$ Monte Carlo repetitions. The parameter values for this data generating process were given by:
\begin{equation*}
	\begin{array}{ll}
		\multicolumn{2}{l}{\beta_2 = (\beta_{20}, \beta_{21}, \beta_{22}, \beta_{23}, \beta_{24}, \beta_{25}, \beta_{26}, \beta_u) = (1, -1, 1, -1, 1, 1, 1, 2),} \\	
		\psi_2 = (\psi_{20},\psi_{21}, \psi_{22}, \psi_{23}) = (-1, 0.5, 0.5, 0.5),  &  \\
		\alpha_1 = (\alpha_{10}, \alpha_{11}, \alpha_{12}, \alpha_{13}) = (0, 1, 1, 2), & \phi_1 = (\phi_{10}, \phi_{11}) = (0, 1),   \\
		\alpha_2 = (\alpha_{20}, \alpha_{21}, \alpha_{22}, \alpha_{23}, \alpha_{24}, \alpha_{25}) = (0, 1, 1, 1, 1, 3), & \phi_2 = (\phi_{10}, \phi_{11}) = (0, -1), \\
		 \sigma_u^2 = \sigma_{x_{11}}^2 = \sigma_{x_{12}}^2  = \sigma_{x_{2}}^2 = \sigma_y^2 = 1, & \varpi = (\varpi_0, \varpi_1, \varpi_2) = (0, 1, 1). \\	
	\end{array}
\end{equation*}
We again varied the posited distributions for the parameters of the unmeasured confounder model and the effect of the unmeasured confounder using the same posited distributions as the single stage simulation study.

\begin{figure}[h]
	\centering
	\includegraphics[width=0.95\linewidth]{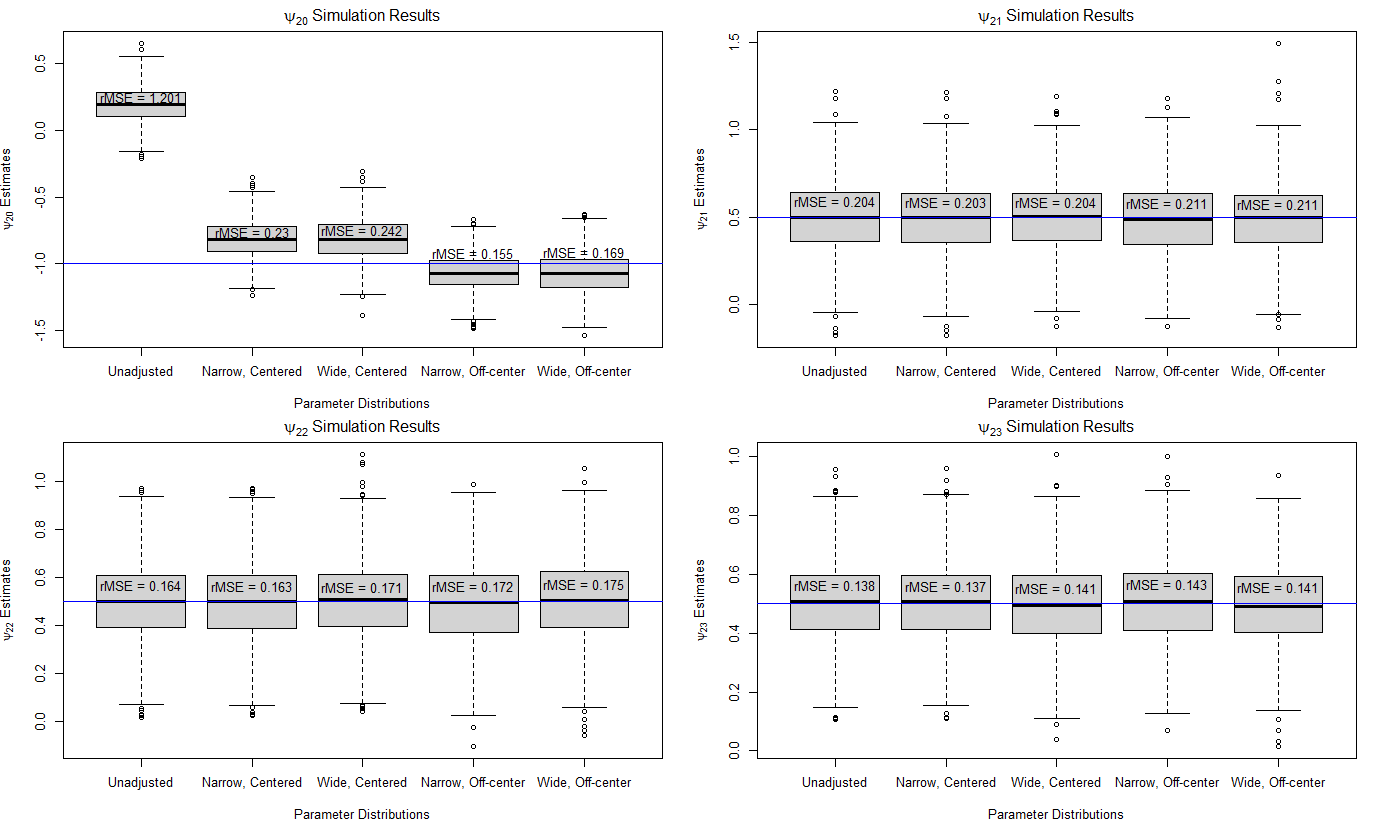}
	\caption{Boxplots of the point estimates for $\psi_{2}$ under an unadjusted model and when using Monte Carlo sensitivity analysis to adjust for bias due to unmeasured confounding for the 2-stage data generating model.}
	\label{fig:psi2_2stage}
\end{figure}

\begin{figure}[h]
	\centering
	\includegraphics[width=0.95\linewidth]{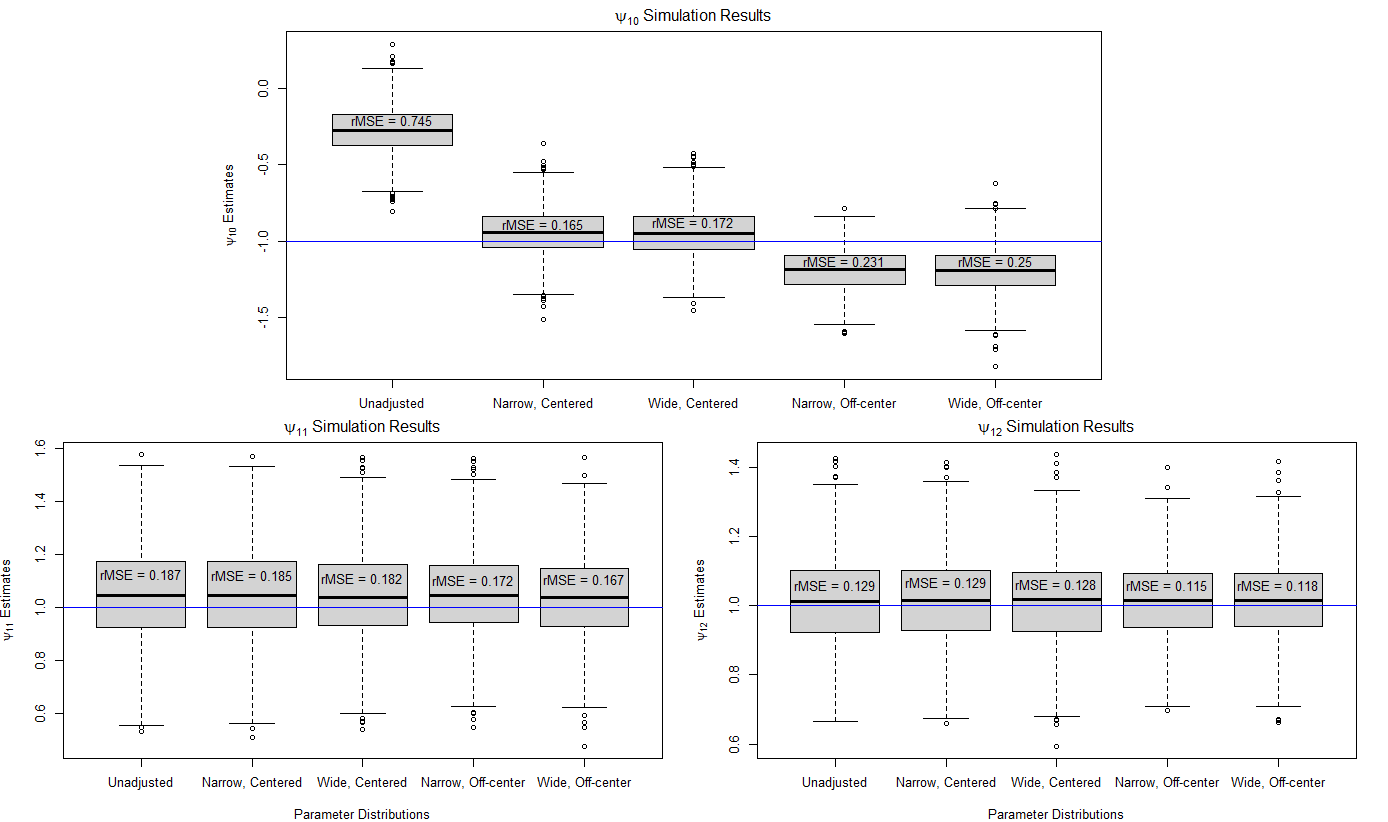}
	\caption{Boxplots of the point estimates for $\psi_{1}$ under an unadjusted model and when using Monte Carlo sensitivity analysis to adjust for bias due to unmeasured confounding for the 2-stage data generating model.}
	\label{fig:psi1_2stage}
\end{figure}

Figures \ref{fig:psi2_2stage} and \ref{fig:psi1_2stage} show boxplots of the point estimates for $\psi_2$ and $\psi_1$, respectively, across the 1000 repetitions. The results for the 2-stage study were very similar to the 1-stage results. In the second stage models, the unmeasured confounder caused significant bias in $\psi_{20}$ in the unadjusted analysis leading to an rMSE of 1.201. Applying our Monte Carlo sensitivity analysis reduced the bias significantly with all scenarios having an rMSE of 0.242 or less. We found that the unmeasured confounder did not cause bias in the estimates of $\psi_{21}$, $\psi_{22}$, and $\psi_{23}$, leading to similar results between the unadjusted and adjusted estimates. There was significant bias in the unadjusted estimate of $\psi_{10}$ for the first stage blip model. The rMSE of the unadjusted estimate is 0.745 while the rMSE of the bias adjusted estimate under the narrow, centered parameter distribution is 0.165. The rMSE increased to 0.25 when using a wide parameter distribution that is not centered on the true parameter value.

As before, we simulated 10000 additional patients and assessed whether the treatment recommended by each of the estimated regimes matched the recommendation of the true optimal regime for each of the 1000 repetitions. Table \ref{propOpt_2stage} displays the proportion of patients who were recommended the same treatment as the true optimal regime when following each of the different estimated regimes. The bias from the unmeasured confounder caused the unadjusted estimated regime to recommend the same treatment as the true optimal regime 81.4\% of the time at the first stage and 70.2\% of the at the second stage. After performing Monte Carlo sensitivity analysis using the narrow, centered posited distribution the proportion increased to 95.1\% at the first stage and 93.7\% at the second stage. The alternative parameter distributions produced similar results.

\begin{table}[h]
	\caption{
		Proportion of patients whose recommended treatment when following each of the estimated regime matches the recommendation of the true optimal regime at each stage for the 2-stage data generating model.
	}
	\label{propOpt_2stage}
	\begin{center}
		\begin{tabular}{|c c c|} \hline
			Parameter Distr. & Stage 1 & Stage 2 \\ \hline
			Unadjusted & 0.814 & 0.702 \\ 
			Narrow, Centered & 0.951 & 0.937 \\ 
			Wide, Centered & 0.949 & 0.936 \\ 
			Narrow, Off-center & 0.951 & 0.948 \\ 
			Wide, Off-center & 0.948 & 0.947 \\  \hline
		\end{tabular}
	\end{center}
\end{table}

We also examined the coverage and width of the confidence intervals generated using Monte Carlo sensitivity analysis for the 2-stage study. Table \ref{cover_2stage_psi2} contains the empirical coverage and width of 95\% confidence intervals for the parameters of the second stage blip model. The empirical coverage of $\psi_{20}$ for the unadjusted analysis was 0\% as it was significantly biased by the unmeasured confounder. Monte Carlo sensitivity analysis produced intervals for $\psi_{20}$ with 100\% coverage for all of the simulation scenarios at the expense of being much wider. The remaining second stage blip parameters, $\psi_{21}$, $\psi_{22}$, and $\psi_{23}$, were not biased in the unadjusted analysis, but the unmeasured confounder led to confidence intervals that did not achieve the nominal coverage probability. Monte Carlo sensitivity analysis produced confidence intervals that achieved the nominal coverage for all sets of posited parameter distributions.

\begin{table}[h]
	\caption{
		Coverage (Cvr.) and average width (Wth.) of the 95\% confidence intervals for $\psi_2$ for the unadjusted analysis and sensitivity analysis under each of the posited parameter distributions for the 2-stage data generating model. * indicates coverages that are significantly different than 95\%.
	}
	\label{cover_2stage_psi2}
	\begin{center}
		\begin{tabular}{|c c c c c|} \hline
			Parameter Distr. & Cvr. (Wth.) $\psi_{20}$ & Cvr. (Wth.) $\psi_{21}$ & Cvr. (Wth.) $\psi_{22}$ & Cvr. (Wth.) $\psi_{23}$  \\ \hline
			Unadjusted & 0.000* (0.339) & 0.810* (0.529) & 0.810* (0.435) & 0.801* (0.359) \\ 
			Narrow, Centered & 1.000* (2.634) & 0.948 (0.803) & 0.953 (0.658) & 0.943 (0.541) \\ 
			Wide, Centered & 1.000* (6.208) & 0.948 (0.804) & 0.951 (0.654) & 0.941 (0.537) \\ 
			Narrow, Off-center & 1.000* (2.775) & 0.932* (0.803) & 0.949 (0.657) & 0.936* (0.542) \\ 
			Wide, Off-center & 1.000* (6.483) & 0.934* (0.805) & 0.937 (0.659) & 0.941 (0.540) \\ 
			\hline
		\end{tabular}
	\end{center}
\end{table}

\begin{table}[h]
	\caption{
		Coverage (Cvr.) and average width (Wth.) of the 95\% confidence intervals for $\psi_1$ for the unadjusted analysis and sensitivity analysis under each of the posited parameter distributions for the 2-stage data generating model. * indicates coverages that are significantly different than 95\%.
	}
	\label{cover_2stage_psi1}
	\begin{center}
		\begin{tabular}{|c c c c|} \hline
			Parameter Distr. & Cvr. (Wth.) $\psi_{10}$ & Cvr. (Wth.) $\psi_{11}$ & Cvr. (Wth.) $\psi_{12}$ \\ \hline
			Unadjusted & 0.001* (0.456) & 0.861* (0.542) & 0.868* (0.382) \\ 
			Narrow, Centered & 1.000* (4.236) & 1.000* (2.609) & 0.989* (2.439) \\ 
			Wide, Centered & 1.000* (7.362) & 1.000* (2.828) & 1.000* (2.590) \\ 
			Narrow, Off-center & 1.000* (4.607) & 1.000* (2.833) & 0.995* (2.663) \\ 
			Wide, Off-center & 1.000* (7.857) & 1.000* (3.081) & 1.000* (2.840) \\ 
			\hline
		\end{tabular}
	\end{center}
\end{table}

Table \ref{cover_2stage_psi1} contains the empirical coverage and average width of 95\% confidence intervals for the parameters of the first stage blip model. The unadjusted confidence interval for $\psi_{10}$ had a coverage of only 0.1\% while the sensitivity analysis resulted in conservative confidence intervals that had coverage of 100\%. For $\psi_{11}$ and $\psi_{12}$, the unadjusted analysis resulted in confidence intervals that did not attain nominal coverage with empirical coverages of 86.1\% and 86.8\%, respectively, despite the unmeasured confounder not resulting in bias in the parameters estimation. The sensitivity analysis intervals were instead conservative with coverages of 100\% and 98.9\% for $\psi_{11}$ and $\psi_{12}$, respectively, for the narrow, centered parameter distributions. The confidence intervals resulting from the sensitivity analysis for the first stage blip model were considerably wider than those for the second-stage blip model parameters.

\section{Application to KPWA EHR Data}
\label{s:ehrs}

We applied the proposed method to data from electronic health records (EHRs). Kaiser Permanente Washington (KPWA) is a health system that provides heath care and insurance to its members. This study used data from EHRs and health insurance claims for all KPWA clients to study the use of antidepressants for treating depression. The data include information on demographics, prescription fills, and depressive symptom severity for 82691 patients that were treated for depression from 2008 through 2018. Severity of depression was assessed through the use of the Patient Health Questionnaire-8 (PHQ) \citep{kroenke_2001}. This is a self-report questionnaire that produces a score ranging from 0 to 24, such that higher values indicate more severe symptoms of depression. The inclusion criteria for the study was that patients must have been 13 years or older, been enrolled in KPWA insurance for at least a year, been diagnosed with a depressive disorder in the year before or 15 days after treatment initiation, had no prescription fills for antidepressant medications in the past year, and had no diagnosis for personality, bipolar, or psychotic disorders in the past year. We additionally required, for this analysis, that patients did not have missing information on obesity, baseline PHQ, or follow-up PHQ at one year after baseline. Our analysis focuses on demonstrating the impact of our sensitivity analysis but does not adjust for missing information, therefore the estimated treatment regimes must be interpreted as potentially biased; any clinical findings should be viewed in this light.

Research has suggested that obesity increases the risk of depression \citep{Luppino_2010}. Weight gain has been found to be a side effect of taking antidepressants \citep{Fava_2000}, with different classes of antidepressants affecting weight differently; treatment guidelines for clinicians recommend using the current weight of a patient as a consideration when prescribing antidepressants \citep{Santasieri_2015}. Obesity is therefore a potentially important confounder which, if unmeasured, could lead to bias in observational studies of their effects, whether average or individualized. We considered a patient to be obese if they had a body mass index (BMI) of 30 or larger. We used this data in two different ways. We assumed obesity was unavailable and estimated a DTR to minimize depression symptoms without adjusting for obesity in the analysis. We then applied the proposed procedure to assess how sensitive the estimated regime was to obesity being unmeasured and compared this to the estimated regime with obesity measured. We also conducted a plasmode simulation study in which we used the real data with simulated obesity and patient outcomes. This allowed us to know the true optimal treatment regime and distribution of obesity.

\subsection{Empirical Evaluation}

Our outcome is the negative of the PHQ score after 1 year so that higher values correspond with better patient outcomes. Equally, we could use the PHQ score and find the regime that minimizes instead of maximizes the mean outcome. The PHQ score after 1 year was given by the PHQ score recorded between 305 and 425 days after initiating treatment that is closest to 365 days after. Patients were treated with one of 17 different antidepressants initially. Selective serotonin reuptake inhibitors (SSRIs) are a class of antidepressants that increase serotonin levels in the brain by decreasing reabsorption by nerve cells and are commonly prescribed due to having generally milder side effects.
We classified the initial treatment received as an antidepressant from either the SSRI class or an alternative class of antidepressants.

We let $Y$ denote the negative of the PHQ score after 1 year. We denote treatment by $A$, such that $A=1$ if assigned an SSRI and $A=0$ if assigned a non-SSRI.
We considered sex, age, baseline PHQ, and obesity as confounders and examined the bias that occurs if obesity was unmeasured. Our outcome model was given by:
\begin{align}
	\label{outcome_model}
	\E(Y|\bX, A;\beta, \psi) = \beta_0 + & \beta_1 \mathrm{SEX} + \beta_2 \mathrm{AGE} + \beta_3 \mathrm{PHQ} + \beta_u \mathrm{OBESE} \\ \nonumber
	& + A(\psi_0 +  \psi_1 \mathrm{SEX} + \psi_2 \mathrm{AGE} + \psi_3 \mathrm{PHQ} )
\end{align}
such that the optimal treatment regime is then given by
recommend an SSRI if $(\psi_0 +  \psi_1 \mathrm{SEX} + \psi_2 \mathrm{AGE} + \psi_3 \mathrm{PHQ})$ is greater than zero.

We found that obesity was correlated with age, baseline PHQ, race, census block education level (EDU), and diagnosis of an anxiety disorder in the prior year (ANX). Race was categorized as Asian, Black or African American, Hispanic, Native Hawaiian/Pacific Islander, American Indian/Native Alaskan, White, other, or unknown and was coded using dummy coding with Asian as the reference group. EDU was given by an indicator for whether less than 25\% of people living in the patient's census block had a college degree. Therefore, to conduct Monte Carlo sensitivity analysis for the bias if obesity is unmeasured, we posited a model for obesity given by:
\begin{align}
	\label{obese_model}
	\Pr(\mathrm{OBESE} = 1| \bX; \zeta) = \Big[ 1 + \exp\{-(\zeta_{0} & + \zeta_{1} \mathrm{AGE} + \zeta_{2} \mathrm{PHQ} +
	\zeta_{3}^T \mathrm{RACE} \\ \nonumber
	& + \zeta_{4}\mathrm{EDU} + \zeta_{5}\mathrm{ANX} + \zeta_{6}A) \} \Big]^{-1} .
\end{align}
We first estimated $\beta_u$ and $\zeta$ using the full data. We then considered obesity to be unmeasured and posited normal distributions for $\beta_u$ and $\zeta$ that are centered at the estimated values of the parameters with standard deviations equal to 0.05 for $\beta_u$, 0.05 for $\zeta_0$, and 0.1 for $\zeta_j$ for $j=1,\dots,6$. The smaller standard deviation for $\zeta_0$ reflects that we have less uncertainty in the prevalence of obesity in our population of interest.
Table \ref{KPWA_coef} contains estimates and confidence intervals for $\psi$ from the full model, the model with obesity unmeasured, and after adjusting for the bias. The adjusted estimates of $\psi$ with obesity unmeasured were close to the estimates from the full model with obesity included. This sensitivity analysis also produced confidence intervals that were wider for the parameters that were biased due to the unmeasured confounding.

\begin{table}[h]
	\caption{
		Estimates and 95\% confidence intervals for treatment decision rule parameters $\psi$ from models with and without obesity and adjusted estimates from the sensitivity analysis for unmeasured confounding of obesity.
	}
	\begin{center}
		\begin{tabular}{|c c c c|} \hline
			Covariate & Full Model & Obesity Unmeasured & Adjusted Est. \\ \hline
			$A$ & 1.59 (-0.53, 3.72) & 1.39 (-0.74, 3.52) & 1.54 (-0.95, 4.17) \\
			$A$  $\times$ SEX & -0.72 (-1.79, 0.34) & -0.69 (-1.76, 0.38) & -0.73 (-2.04, 0.59) \\
			$A$ $\times$ AGE & 0.00 (-0.03, 0.03) & 0.00 
			(-0.03, 0.03) & 0.00 (-0.04, 0.04)\\
			$A$ $\times$ PHQ & -0.12 (-0.22, -0.02) & -0.12 (-0.22, -0.02) & -0.12 (-0.24, -0.01) \\ \hline
		\end{tabular}
	\end{center}
	\label{KPWA_coef}
\end{table}

\subsection{Plasmode Simulations}

We also conducted a plasmode simulation study in which we used the real data with simulated obesity and patient outcomes. This allowed us to know the true optimal treatment regime and unmeasured confounder distribution. The data were simulated using models \ref{outcome_model}  and \ref{obese_model} that we used for the real data analysis with the outcome having a random error given by $\epsilon \sim N(0, 1)$.

We let the value of $\beta$, $\psi$, and $\zeta$ be given by the estimated values from the data which can be found in Section C of the supplementary material. For conducting the sensitivity analysis, we posited the same normal distributions for $\beta_u$ and $\zeta$ that we used previously for the real data example. We simulated 1000 data sets and estimated the adjusted parameter estimates for each data set. Table \ref{KPWA_plasmode} contains the estimates of $\psi$ averaged over the 1000 simulated data sets in addition to the empirical coverage and average width of 95\% confidence intervals for $\psi$. The unadjusted estimate for $\psi_0$ was biased with an average estimate of 1.47 as opposed to the true value of 1.59. Adjusting for obesity reduced the bias and resulted in an estimate of 1.54. The 95\% confidence intervals from the unadjusted analysis did not achieve the nominal rate for any of the parameters. Monte Carlo sensitivity analysis resulted in slightly wider intervals that had the correct coverage.

\begin{table}[h]
	\caption{
		Estimates (Est.), coverage (Cvr.), and average width (Wth.) of 95\% confidence intervals for treatment decision rule parameters $\psi$ after adjusting for the unmeasured confounder obesity averaged over 1000 plasmode data sets using data from Kaiser Permanente Washingtion. * indicates coverages that are significantly different than 95\%.
	}
	\begin{center}
		\begin{tabular}{|c c c c c|} \hline
			& Est. $\psi_{0}$ & Est. $\psi_{1}$ & Est. $\psi_{2}$ & Est. $\psi_{3}$ \\ \hline
			True Value & 1.59 & -0.72 & 0.00 & -0.12 \\
			Unadjusted & 1.47 & -0.71 & 0.00 & -0.12 \\
			Adjusted & 1.54 & -0.73 & 0.00 & -0.12 \\ \hline
			& Cvr. (Wth.) $\psi_{0}$ & Cvr. (Wth.) $\psi_{1}$ & Cvr. (Wth.) $\psi_{2}$ & Cvr. (Wth.) $\psi_{3}$ \\ \hline
			Unadjusted & 0.81* (0.91) & 0.87* (0.46) & 0.85* (0.01) & 0.86* (0.04) \\
			Adjusted & 0.95 (1.19) & 0.94 (0.57) & 0.94 (0.02) & 0.94 (0.05) \\ \hline
		\end{tabular} 
	\end{center}
	\label{KPWA_plasmode}
\end{table}

\section{Discussion}
\label{s:discussion}

We proposed a method for conducting sensitivity analysis for bias due to unmeasured confounding in the estimation of DTRs. We used Monte Carlo sensitivity analysis to estimate a bias adjusted treatment regime and construct confidence intervals for the parameters indexing the optimal regime that account for the uncertainty due to unmeasured confounding. This procedure is straightforward to implement and can be used for continuous or binary unmeasured confounders. This approach was found to perform well for both simulated and real data.

This method requires positing parametric models for the relationship between the unmeasured confounder(s) and the outcome, treatment, and measured confounders. Moreover, the model for the conditional mean of the unmeasured confounder needs to be of high quality, which can be challenging without external data or domain expertise. We also must posit probability distributions for the parameters indexing these models. These distributions do not need to be centered on the true value of the parameter to construct confidence intervals, they only need the true value to not be in the tails of the distribution. If there are high levels of uncertainty about these parameters, wider distributions can be posited resulting in wider confidence intervals for the parameters indexing the estimated optimal treatment regime. We posited normal distributions for these parameters, but any probability distribution can be used that accurately reflect the prior information about the true value of these parameters. Here we focused on treatment regimes that are estimated using dWOLS, though this approach can be applied to any regression-based method of estimation. Direct-search or value-search estimators are an alternative approach to estimating DTRs that our proposed method could not be directly applied to \citep{Orellana10, laber_2015}. We leave extensions to direct-search estimators as future work.

\section*{Acknowledgments}

This work was supported by the National Institute of Mental
Health of the National Institutes of Health under Award Number R01 MH114873. The content is solely the responsibility of the authors and does not necessarily represent the official views of the National Institutes of Health. EEMM is a Canada Research Chair (Tier 1) and acknowledges the support of a chercheur de m\'erite career award from the Fonds de Recherche du Québec, Santé.

\bibliographystyle{Chicago}
\bibliography{sensitivity_dWOLS}

\end{document}